**Formation and Magnetic Drive of 1D-Confined Topological Defect Lines in Crystals with $D_{2d}$ Symmetry**

Bimalesh Giri[1,2], Filipp N. Rybakov[3], Nikolai S. Kiselev[4], Tamer Karaman[5], Satya Prakash Senapati[1,6], Ratiranjan Samal[1], Dola Chakrabartty[1,5], Anshuman Padhi[2], Abhay Kant Srivastava[2], Banabir Pal[2], Amir H. Tavabi[6], Felix Bu¨ttner[5,7], Istva´n Ke´zsma´rki[5], Rafal E. Dunin-Borkowski[7], Stuart S. P. Parkin[2], and Ajaya K. Nayak[1,8, *]

[1]School of Physical Sciences, National Institute of Science Education and Research, An OCC of Homi Bhabha National Institute, Jatni-752050, India

[2]Max Planck Institute of Microstructure Physics, Weinberg 2, 06120 Halle (Saale), Germany

[3]Department of Physics and Astronomy, Uppsala UniversityBox-516, Uppsala SE-751 20, Sweden

[4]Peter Grünberg Institute, Forschungszentrum Jülich and JARA, 52425 Jülich, Germany

[5]Experimental Physics V, Center for Electronic Correlations and Magnetism, Institute for Physics, University of Augsburg, 86159 Augsburg, Germany

[6]Ernst Ruska-Centre for Microscopy and Spectroscopy with Electrons, Forschungszentrum Jülich, 52425 Jülich, Germany

[7]Helmholtz-Zentrum Berlin, 14109 Berlin, Germany

[8]Center for Interdisciplinary Sciences (CIS), National Institute of Science Education and Research, Jatni, 752050 Bhubaneswar, India

[*]Corresponding author: ajaya@niser.ac.in

## Abstract

Designing and controlling topological spin textures within non-collinear magnetic backgrounds is central to the development of low-dissipation spintronic devices. Here we propose a general mechanism—based on tunable magnetic anisotropy—that enables the realization of one-dimensional topological defect lines embedded within intrinsic spin-spiral states of uniaxial $D_{2d}$-symmetric magnets, and is broadly applicable across diverse material platforms. Using Lorentz transmission electron microscopy and off-axis electron holography, we show that introducing in-plane magnetic anisotropy into the helical ground state stabilizes adjacent domains of conical helices with in-plane propagation vectors, sharing the same handedness but exhibit opposite net magnetization. The resulting interfacial mismatch generates confined bimeron-chains. Significantly, these defect lines are not merely induced merons; their presence is mandated by the specific $D_{2d}$ symmetry constraints inherent to the system. Applying weak magnetic fields along the conical axis drives counter-propagating motion of adjacent bimeron-chains along the helical stripes, while anisotropy-mediated confinement enables track-stable transverse drift free transport. The present work establishes a clear route to stabilize and control topological spin textures within intrinsically non-collinear magnetic states, advancing fundamental understanding and enabling low-dissipation, scalable spintronic device concepts.

The concept of topological defects[1] in condensed matter physics offers key insights into the diverse physical properties and emergent behaviors of condensed matter systems[2-6]. Domain walls (DW) viewed as one-dimensional defect lines separates regions having different order parameter in ferroelectric and ferromagnetic (FM) materials have attracted considerable attention in the field of nanoelectronics and spintronics[7-9]. In conventional collinear ferromagnets, DWs are typically classified as Bloch[10] or Néel types[11]; however, under appropriate magnetic interactions they can accommodate a variety of topological defects, such as Bloch lines, (anti)merons, bimerons, and skyrmions etc[12-17]. While domain walls in collinear ferromagnets are well understood, their counterparts in non-collinear magnetic systems remain largely unexplored, representing a major gap in our understanding of complex magnetic textures. Crucially, we report that within the specific $D_{2d}$ system under investigation, symmetry constraints mandate the unavoidable presence of bimerons within the DW structure formed at the interface of conical-helix domains with opposite net magnetization- a defining and fundamental characteristic not always present or guaranteed across previously studied systems.

Among the diverse classes of non-collinear magnetic textures, spin spirals provide a prototypical model system for investigating domain-wall physics. For example, helical and cycloidal states are prototypical non-collinear magnetic configurations characterized by a periodic rotation of the magnetization along a propagation vector ($\boldsymbol{q}$) [18-25] which is very similar to stripe domain states[26-28]. In chiral or non-centrosymmetric materials where Dzyaloshinskii–Moriya interaction (DMI) is present such modulated states often appear as ground states, from which more complex spin textures can emerge upon external perturbation[22,23,29-32]. Furthermore, the inherent modulated character of these spin-spiral states allows them not only to be precursors for complex phases such as skyrmion lattices[29-37]but also to host stable two-dimensional topological defects. Even, the DWs between helical domains with different $\boldsymbol{q}$ vector direction can acquire non-trivial topology depending on the relative phase and propagation direction of the underlying spin spirals[38,39]. In this context, interfaces between distinct spin-spiral states offer a natural platform for localization, as the mismatch of their underlying modulations can give rise to non-trivial magnetic configurations. In this work, we exploit a model system with $D_{2d}$ symmetry that exhibits asymmetric DMI to engineer the emergence of spatially localized bimeron-chains. The $D_{2d}$ symmetry systems naturally host single-domain helical magnetic states with opposite chiralities, whose $\boldsymbol{q}$ vectors are oriented perpendicular to each other due to anisotropic DMI[40-43]. However, in the systems under study, additional spontaneous symmetry breaking is possible. Namely, easy-plane anisotropy can modify the planar spin helix into cone-like ones in two ways, differing in opposite net magnetizations. This, in turn, enables the emergence, of 1D-confined bimeron chains, which serve as topological defect lines separating domains with two types of such spirals. Usually, spin-conics state appears when the magnetic field is applied along the propagation axis. In contrast, here we propose a design strategy that stabilizes spin conics in the $D_{2d}$ system by reducing the uniaxial magnetic anisotropy (UMA) along the crystallographic *c*-axis ([001] direction) so much that the effective anisotropy (including dipolar effects) becomes easy-plane-type, which we also call in-plane (IP) type. This is analogous to switching the effective perpendicular magnetic anisotropy (PMA) in thin films from positive to negative. Note that both UMA and IP anisotropy are consistent with the point symmetry group $D_{2d}$. Figure 1a (left panel) shows the schematic of a right-handed helix, where spins, rotating in the plane perpendicular to the helical (*Y)* axis, trace out a right-handed screw. Let’s consider this helical state as the ground-state magnetic configuration in the UMA or isotropic system possessing DMI. Introducing a sufficient magneto crystalline or shape induced IP anisotropy into the system causes a tilting of all of the magnetic moments towards the helical axis, thereby giving rise to two conical right-handed helices, which are distinguished by their net magnetization pointing along opposite directions (as illustrated in the middle and right panels of Fig. 1a). In this scenario, a small uniform magnetization emerges along the helical axis. The two possible choices of the net magnetization, lead to two energetically degenerate domains, namely conical states with the same handedness but opposite net magnetization.

The anisotropic DMI imposed by $D_{2d}$ symmetry selects the [100] and [010] directions within the basal plane as the energetically preferred propagation directions of the helical states. In Mn-based $D_{2d}$ systems, compositional tuning can modify the coupling between the two magnetic sublattices and redistribute the magnetization between its in-plane and out-of-plane components[44,45]. This provides a route to transform the helical state into two degenerate conical-spiral states with opposite net magnetizations, as illustrated schematically in Fig. 1**a,b**. To test this mechanism, we performed micromagnetic simulations in the *Excalibur* code[46] for a $D_{2d}$-symmetric magnetic plate using the following energy functional:

$$E = \int_{V_m} \left(A \sum_{i=x,y,z} |\nabla m_i|^2 + D \left[\Lambda_{zy}^{(x)} + \Lambda_{zx}^{(y)}\right] - Ms\, \boldsymbol{m}.\boldsymbol{B}\right) \mathrm{dr} + \frac{1}{2\mu_0} \int_{R^3} \left(\sum_{i=x,y,z} |\nabla A_{d,i}|^2\right) dr$$

Here, A is the exchange stiffness, and D is the anisotropic DMI constant. The magnetic field B includes both the external and demagnetizing fields.

Where $\mathbf{m}(\mathbf{r}) = \mathbf{M}(\mathbf{r})/M_s$ is a unit vector field that defines the direction of the magnetization, $M_s = |M(r)|$ is the saturation magnetization and $\mu_0$ is the vacuum permeability. The magnetic field $\mathbf{B} = \mathbf{B}_{ext} + \boldsymbol{\nabla}\times\mathbf{A}_d$, is the sum of the external magnetic field and the demagnetizing field, where $\mathbf{A}_d(\mathbf{r})$ is the component of the magnetic vector potential induced by the sample. Further details of the numerical implementation and the simulation parameters are provided in the Methods.

Figures 1**c-e** show stable configurations found by direct energy minimization. For a sample having a shape in the form of a plate, the demagnetizing field results in an effective easy -plane anisotropy. The lateral size of the simulated domain is $1\mu$m×$1\mu$m. For the case of an extended plate tends to the value $\kappa = \frac{1}{2}\mu_0 M_s^2 \approx 2.5\ kJ/m^3$. As a result, for chosen set of parameters, the lowest-energy state of the extended plate is a conical-helix with an in-plane ***q***-vector. It is worth noting that due to the presence of easy plane anisotropy, the spins do not trace flat circles on the spin sphere as in the case of ordinary conical spirals. Similar configurations have been studied earlier[47]. As illustrated in Fig. 1,

the conical spirals with opposite net magnetization can form multi-domain states. The boundary between such domains represents a spontaneously formed chain of bimerons that corresponds to the lowest-energy configuration. The configurations depicted in Fig. 1**c**- **e** were obtained by energy minimization using an initial state composed of two domains with conical phases, having opposite net magnetizations, separated by a sharp domain boundary.

Figure 1**c** shows an alternating chain of merons with topological indices (0, −1) and (1, 0), which emerges within a domain wall in the quasi 2D limit (plate thickness 5 nm). As indicated in the inset, the total topological index of an individual meron pair, or bimeron, is (1, −1). The emergence of bimerons in the system with easy plane anisotropy is fully consistent with the homotopy-group analysis provided in Ref.[48]. The classifying group of merons in such systems is $\pi_2(s^2, s^2 \setminus P_1, P_2\}) = \mathbb{Z} \times \mathbb{Z}$, where $P_1$ and $P_2$ are two opposite hard directions in the case of easy plane anisotropy. The corresponding topological invariant represents an ordered pair of integers ($q_t$, $q_b$), where $q_t$ and $q_b$ stand for the so-called *top* and *bottom* topological charges. For details, see Ref. [48] and Section 1 of the Supplementary Materials. Figures 1**d** and **e** illustrate analogous configurations in a thicker plate with a thickness of 185 nm. The differences between the configurations in Fig. 1**d** and **e** are the orientation of the net magnetization in the left and right domains and the sign of the topological indices. As a consequence, the theoretical Lorentz TEM contrast, shown below the simulated domains for the two cases, differs markedly. In particular, the characteristic contrast of the bimeron-chains displayed in the lower insets of Fig. 1**d** and **e** is fully inverted when switching between overfocused and under focused imaging conditions.

Supplementary Video 1 presents interactive real-time simulations of bimerons and the calculation of the corresponding topological indices. This part of the simulations is carried out with switched off dipole-dipole interactions in order to demonstrate the fundamental nature of the phenomena resulting from the minimal model. The video also demonstrates bimeron annihilation induced by a weak in-plane magnetic field applied along the net magnetization in one of the domains. This annihilation process corresponds to a null homotopic fusion of merons carrying opposite topological charges, such as (−1, 0) and (+1, 0) or (0, −1) and (0, +1). Observation of the theoretically predicted contrast, together with the characteristic meron annihilation process, provides a clear experimental signature for identifying bimerons.

Motivated by our simulations, we present the first experimental observation of 1D-confined bimeron-chains as topological defect lines within a natural one-dimensional conical-helix background. To achieve our goal, we substitute Sn with indium (In) in the antiskyrmion host Mn-Pt(Pd)-Sn, an inverse tetragonal Heusler system with $D_{2d}$ symmetry[40]. The $D_{2d}$ symmetry restricts the DM vectors within the tetragonal basal plane with $D_x$= - $D_y$, where $D_x$, $D_y$ are components of the DM vector along the mutually orthogonal *a* and *b* crystallographic directions, respectively. We earlier showed that replacing Sn by In results in a non-collinear magnetic state in $Mn_{1.5}PtIn$ with an uncompensated magnetic moment of ∼1.6 $\mu_B$/f.u. [45]. Moreover, a comparison between the low temperature neutron diffraction data of $Mn_{1.4}PtSn$ and $Mn_{1.5}PtIn$ indicates that a large magnetization can be induced in the basal plane with In substitution. Accordingly, to realize a tunable redistribution of magnetic moments between the IP and OP orientations, we systematically substitute In with Sn in the $D_{2d}$-symmetric $Mn_{1.4}Pt(Pd)Sn_{1-x}In_x$ system [please refer to **Supplementary Section-2** for structural details]. The ground state magnetic configurations obtained from the neutron diffraction study for $Mn_{1.4}PtSn_{1-x}In_x$ with *x* =0.2, 0.4 and 0.6 are presented in Fig. 2**a-c**. Here the Mn atoms occupy two different magnetic sublattices. For *x* = 0.2, the Mn moments within both magnetic sublattices align along the *c*-axis above the spin-reorientation transition temperature $T_{SR}$, resulting in an OP magnetization component $M_Z$ [Fig. 2**a**]. In general, this type of spin arrangement favors the antiskyrmion phase[40]. In the case of *x* = 0.4, the Mn moments within both magnetic sublattices align at a certain angle with respect to the *c* axis. Most importantly, the IP component of the magnetic moments of the Mn atoms sitting on the 4a and 8d Wyckoff positions exhibit an AFM alignment, whereas the OP components still show FM ordering. Interestingly, for *x* = 0.4, both magnetization components are comparable in magnitude, i.e., $M_Z \sim M_Y$ [Fig. 2**b**]. For $x$ = 0.6, the Mn moments at the 8d position align completely in the IP direction [Fig. 2**c**]. For the detailed analysis of magnetic and neutron diffraction results, please refer to **Supplementary** Section-3 and Section-4.

A detailed variation of the ratio of $M_Y$ /$M_Z$ with temperature and different In concentrations is plotted in [Fig. 2**d**]. The region with $M_Y \approx M_Z$ is of particular interest, as a further increase in In concentration leads to dominant IP anisotropy in the system. The variation of effective magnetic anisotropy ($K_{eff}$) and saturation magnetization ($M_S$) with In concentration for the $Mn_{1.4}PtSn_{1-x}In_x$ system is shown in Fig. 2(e) [please refer to **Supplementary Section-3** for the calculations of $K_{eff}$]. Although $M_S$ monotonically decreases with $x$, $K_{eff}$ initially decreases up to $x$ = 0.4 and then increases at higher In concentrations, resulting in a local minimum at $x$ = 0.4, which is the sample of particular interest in this study. However, the grain size of $Mn_{1.4}PtSn_{0.6}In_{0.4}$ is found to be too small to prepare a single-crystalline lamella for subsequent studies. Therefore, we doped a small amount of Pd for Pt, which significantly enhances the grain size in $Mn_{1.4}Pt_{0.9}Pd_{0.1}Sn_{1-x}In_x$ samples. As shown in Fig. 2(f), this small Pd doping does not affect the behavior of $M_S$ and $K_{eff}$, so that a local minimum is still observed for $x$ = 0.4. To exactly determine the anisotropy direction for $Mn_{1.4}Pt_{0.9}Pd_{0.1}Sn_{0.6}In_{0.4}$, we have performed direction dependent Hall resistivity measurements using single-crystalline devices fabricated using Focused Ion Beam (FIB) [Fig. 2(g)]. It is found that at $T$ =150 K, the Hall resistivity curves for the IP ($H\perp$ c) follows exactly that of the OP ($H\parallel$ c) orientation, indicating the presence of equal magnitudes of IP and OP anisotropies. Below $T_{SR}$, i.e., for $T\leq$ 100 K, the Hall resistivity for the IP ($H\perp$ c) configuration steeply increases and saturates faster than the OP ($H\parallel$ c) orientation, demonstrating easy-plane anisotropy at these temperatures [**Supplementary Section-5**]. Therefore, we expect a transition in the observed magnetic ground state to occur below $T$ =150 K.

To directly visualize topological spin textures in real space, we have performed a detailed Lorentz transmission electron microscopy (LTEM) study on thin lamellae of the $Mn_{1.4}Pt_{0.9}Pd_{0.1}Sn_{0.6}In_{0.4}$ sample [**Supplementary Section-6**]. Figure 3**a-e** show under focused (UF) LTEM images recorded from [001] oriented thin plates with a thickness of approximately 150 nm. When the lamella is cooled below $T_C$, helical-like magnetic structure propagating along the [010] direction appear as the zero-field ground state and persist down to $T$ = 123 K, as shown in Fig. 3**a**. On further cooling to $T$ = 118 K, weaker bright and dark contrasts appear superimposed on the helical-like magnetic stripes, as indicated by arrows in Fig. 3**b**. These additional magnetic contrasts become more pronounced at $T$ = 113 K, forming alternating bright and dark chain-like patterns (Fig. 3**c**). The intensity of these chain-like magnetic patterns increases progressively with decreasing temperature, as observed at $T$ = 113 K (Fig. 3**d**) and $T$ = 93 (Fig. 3**e**). At $T$ = 93 K, an over focused (OF) LTEM image is also recorded (Fig. 3**e**), revealing a reversal of contrasts of these chain-like patterns. Notably, these alternating bright and dark chains align along the [010] direction, which coincides with the axis of the helical spirals in this compound. As demonstrated by our simulations in Fig. 1**d & e**, the observed chain-like features are likely attributable to the emergence of bimerons at the interfaces of conical spirals with the same handedness but opposite in-plane uniform magnetization.

To substantiate our conjecture, we have carried out a transport-of-intensity equation (TIE) analysis of the LTEM images acquired at $T$ = 93 K [**Supplementary Section-7**]. As can be seen from Fig. 3**g**, the TIE maps of the LTEM images clearly establishes the presence of two oppositely magnetized IP domains separated by a domain boundary consisting of bimeron-chains. The domains marked with 1 & 3 display a magnetization profile that is distinct from that observed in domains 2 & 4. The emergence of these domain-like patterns is associated with the spatial variation of the in-plane magnetization component superimposed on the background of a helical structure. A closer examination of domains 1 and 3 shows that the in-plane magnetization, turning the helical state into a conical one, points along the (010) axis, while for domains 2 and 4 it is reversed. These alternating structures are well resolved in Fig. 3**g**. The boundary between these conical helices with opposite in-plane magnetizations gives rise to a bimeron magnetic profile. As illustrated in Fig. 3**h & i**, the topological charge of the bimerons between domains 2 & 3 is opposite to that between domains 3 & 4. The observed spin textures are fully consistent with our simulated results presented in Fig. 1**d & e**. It can also be seen that the TIE map of the magnetic stripe patterns at 173 K reveals a standard helical like configuration without any modulation of the local magnetic field (Fig. 3**j**).

The conical-helix and interfacial bimeron-chains observed in our LTEM experiments are further validated by off-axis electron holography measurements, as shown in Fig. 3**k**. The IP magnetic induction map exhibits distinct different features that are markedly different from those usually associated with a helical magnetic state[49]. In contrast to the helical magnetic state, where field lines propagate exclusively along the magnetic stripe, the present magnetic induction map displays a wave-like behavior, with field lines containing components both along and perpendicular to the magnetic stripes. Analogous to the TIE mapping of the LTEM image, the magnetic induction map reveals domain-like features, where conical helices in domains 1 & 3 have in-plane magnetization opposite to those in domains 2 & 4. In domains 2 & 4, a closer examination of the interface between the two magnetic stripes reveals a blue contrast, corresponding to the magnetic field oriented along the [0-10] direction. By the same analysis, domains 1 & 3 exhibit a conical helix with a net magnetic field along [010]. At the interfaces between domains 2 & 3 and 3 & 4, closed field lines with reversed polarity are observed, consistent with bimeron-chains exhibiting opposite topological charge. These experimental results confirm the formation of bimeron-chains at the interface of in-plane conical helices of opposite net magnetization.

Next, we investigate the field evolution of the bimeron-chains under a small in-plane (IP) magnetic field. First, the sample is tilted by about 4° along the [010] direction from the zone axis to introduce a small field component along the conical-helix propagation direction (Fig. 4**a-d).** We identify five distinct conical-helix domains, with the direction of the net staggered magnetization indicated by the white arrows. Adjacent domains possess oppositely oriented net staggered magnetic moments. Upon tilting the sample to introduce a small in-plane magnetic field along the [010] direction, domains 2, 4, and 5 expand, whereas domains 1 and 3 shrink (Fig. 4**b**). As the magnetic field is further increased, two pairs of bimeron chains with opposite topological charges annihilate at the interfaces between domains 2 and 3, and between domains 3 and 4 (Fig. 4**c**). Increasing the field further to H = 0.538 T leads to the annihilation of the remaining pair of bimeron chains, resulting in an almost single-domain conical-helix state (Fig. 4**d**). The lateral motion of the bimeron chains is driven primarily by the in-plane component of the applied magnetic field. The two conical-helix domains possess opposite net in-plane magnetizations along the helical axis. Consequently, the domain whose net magnetization is parallel to the in-plane field expands, whereas the domain with antiparallel magnetization shrinks. The out-of-plane field component could, in principle, modify the conical spin structure. However, our experimental observations indicate that, within the investigated field range, the conical-helix state remains stable. The primary role of the out-of-plane field is therefore to gradually evolve the conical helix to a field-polarized phase at higher fields rather than to drive the lateral motion of the bimeron chains.

To further confirm the role of the in-plane magnetic field in manipulating the conical-helix domains and driving the motion of the bimeron chains, we performed additional experiments on a separate sample using a small total magnetic field of only 0.025 T (Fig. 4**e-p**). After cooling the sample to 93 K, an out-of-plane magnetic field of 0.025 T was first applied at zero tilt (Fig. 4**e**). No noticeable change in the magnetic configuration was observed under this condition. The sample was then tilted to introduce an in-plane magnetic field component. At α = -6.6°, domains 2 and 4 expand, whereas domains 1, 3, and 5 contract (Fig. 4**f**). With increasing tilt angle, the two outer bimeron chains are successively annihilated at the sample edge at α = -11.5° (lower chain, Fig. 4**g**) and α = -16.2° (upper chain, Fig. 4**h**). As a result, only a single pair of bimeron chains with opposite topological charges remains, as highlighted by the dotted box in Fig. 4**h**. The tilt angle was then reduced to α = -1.6°, increasing the separation between the two remaining bimeron chains (Fig. 4**i**). By subsequently increasing the tilt angle to α = +11°, we selectively annihilate one of the two chains, thereby stabilizing an isolated single bimeron chain, as shown in the dotted box in Fig. 4**i**. Controlled back-and-forth motion of this single bimeron chain is achieved by further varying the tilt angle (Fig. 4**k** and **l**). The corresponding magnetization directions of the conical-helix domains at different tilt angles are revealed by the TIE-reconstructed images shown in Fig. 4**m–p**. These results demonstrate that the in-plane magnetic field not only enables deterministic manipulation of the conical-helix domains and bimeron-chain motion but also provides a route to stabilize an isolated single bimeron chain. This capability is particularly important from an application perspective, as it establishes a practical pathway for realizing a single bimeron chain that can be driven along the helical magnetic stripe with intrinsically transverse deflection free motion.

The experimental results are in good agreement with our micromagnetic simulations that show annihilation of bimeron pairs with opposite topological charge by the application of IP magnetic field (Supplementary Video 1). Most importantly, the drift of the bimerons is constrained strictly along the stripe direction. This observation clearly mimic the motion of the confined magnetic objects occurs exclusively along the stripes. Such a motion observed in the present work does not arise from cancellation of oppositely directed gyrocoupling vectors across the specimen thickness or sublattices, as occurs in antiferromagnetic skyrmions[50,51]. In antiferromagnetic skyrmion systems, two skyrmions with opposite polarities reside on different magnetic sublattices, leading to cancellation of the gyrotropic (Magnus) force and, consequently, vanishing skyrmion Hall effect[50,51]. In our case, the bimerons are not localized on the collinear background, which is an important condition for the Thiele formalism. Rather, the magnetic objects embedded in helical background experience a confining potential barrier that counteracts the Magnus force, causing them to move along the stripes[52]. Although a few studies have reported the coexistence of field-induced skyrmions with broken helical stripes[53,54], the present study reports the first of its kind regarding confined topological objects within a truly helical background.

The $D_{2d}$ symmetry permits two energetically equivalent conical-like states with identical chirality but opposite net magnetization, such that the system naturally separates into degenerate domains during cooling. The interface between these domains cannot be described by a conventional domain wall. We emphasize that (bi)merons presence is mandated here making our findings drastically different to other systems where (bi)merons are possible excitations on top of the domain wall[13-17,26]. Figure 5 illustrates geometrically the reason for the above phenomena. In the case of ordinary domains (Fig. 5**a**), the domain wall is essentially a redirection of the magnetization vector, and its minimal model can be represented by a curve on the spin sphere connecting the left and right values of magnetization. At the same time, there is the possibility of stabilizing more complex walls, including those containing merons, such as a cross-tie domain wall[26]. In contrast, for the system under study, the set of magnetization values inside the wall must fill the surface on the spin sphere between the two loops (Fig. 5**b**). Such a fractionalization of the spin sphere is the essence of merons, making them a necessary component of the domain wall. Therefore, the formation of the bimeron chains is an intrinsic consequence of the symmetry and does not require strain, explicit symmetry breaking, or defect-induced pinning associated with grain boundaries.

Most importantly, the current approach of tuning the system's anisotropy in D2d-symmetric materials to realize thermodynamically stable 1D topological defect lines provides a foundational framework for achieving localized topological objects in other systems with analogous symmetry. For example, $S_4$-symmetric systems, which are also known to host room-temperature antiskyrmions[55] demonstrate the viability of this symmetry class beyond conventional $D_{2d}$ Heusler compounds. Furthermore, magnetic anisotropy in these systems can be effectively tuned via compositional engineering, enabling controlled stabilization of desired magnetic phases[56]. In this context, it is worth noting that the formation of a two-dimensional (anti)meron lattice in the β-Mn-type cubic chiral-lattice magnet $Co_8Zn_9Mn_3$, stabilized by in-plane magnetic anisotropy[57], has been reported as a metastable configuration, in contrast to the naturally occurring bimeron states observed in the present system. Consequently, such materials are expected to attract considerable attention for analogous investigations into tunable and localized topological spin textures. For example, the bimeron-chains with opposite topological charge in the present $D_{2d}$-symmetric system provide an excellent playground for non-reciprocal transport effects, such as the electrical magnetochiral dichroism[58,59]. Due to the coexistence of chirality and net magnetization along the helical axis, the system is expected to exhibit different resistivities for currents flowing in opposite directions along the helical axis. On the one hand, this can provide a tool to electrically distinguish stripe domains hosting conical helices with the same handedness but opposite net magnetizations. The inverse effect can be used to manipulate these stripes, i.e., extend one type and squeeze the other by the injection of uniform currents parallel to the helical axis. Therefore, this approach also provides an electrical mean to control the observed bimeron-chains.

## Conclusion

Our study demonstrates a robust strategy to stabilize and control confined topological magnetic states in magnetic crystals with $D_{2d}$-symmetry through anisotropy engineering. By creating oppositely magnetized domains consisting in-plane conical helices with net opposite magnetization, we show that their interfaces are bimeron chains characterized by topological charges. Such 1D defect lines can be fixed in place or driven in a controlled manner along the stripe axis. This anisotropy-mediated confinement not only suppresses transverse motion but also enables directional manipulation of the topological defects. Our findings establish a versatile platform for engineering anisotropy-controlled spin-spiral landscapes and confined topological quasiparticles across bulk crystals and thin-film heterostructures, enabling further exploration of their driven dynamics and guiding the design of next-generation spintronic devices based on robust, controllable topological states.

## References:

1 Mermin, N. D. The topological theory of defects in ordered media. *Rev. Mod. Phys.* **51**, 591-648 (1979).

2 Sondhi, S. L., Karlhede, A., Kivelson, S. A. & Rezayi, E. H. Skyrmions and the crossover from the integer to fractional quantum Hall effect at small Zeeman energies. *Phys. Rev. B.* **47**, 16419-16426 (1993).

3 Neubauer, A. *et al.* Topological Hall Effect in the *A* Phase of MnSi. *Phys. Rev. Lett.* **102**, 186602 (2009).

4 Balke, N. *et al.* Enhanced electric conductivity at ferroelectric vortex cores in BiFeO3. *Nat. Phys.* **8**, 81-88 (2012).

5 Nagaosa, N. & Tokura, Y. Topological properties and dynamics of magnetic skyrmions. *Nat. Nanotechnol.* **8**, 899-911 (2013).

6 Yang, K. *et al.* Wigner solids of domain wall skyrmions. *Nat. Commun.* **12**, 6006 (2021).

7 Allwood, D. A. *et al.* Magnetic Domain-Wall Logic. *Science* **309**, 1688-1692 (2005).

8 Parkin, S. S., Hayashi, M. & Thomas, L. Magnetic domain-wall racetrack memory. *Science* **320**, 190-194 (2008).

9 Catalan, G., Seidel, J., Ramesh, R. & Scott, J. F. Domain wall nanoelectronics. *Rev. Mod. Phys.* **84**, 119-156 (2012).

10 Bloch, F. Zur Theorie des Austauschproblems und der Remanenzerscheinung der Ferromagnetika. *Zeitschrift für Physik* **74**, 295-335 (1932).

11 Néel, M. L. Propriétés magnétiques des ferrites ; ferrimagnétisme et antiferromagnétisme. *Ann. Phys.* **12**, 137-198 (1948).

12 Malozemoff, A. P. & Slonczewski, J. C. Effect of Bloch Lines on Magnetic Domain-Wall Mobility. *Phys. Rev. Lett.* **29**, 952-955 (1972).

13 Gao, Y. *et al.* Spontaneous (Anti)meron Chains in the Domain Walls of van der Waals Ferromagnetic $Fe_{5-x}GeTe_2$. *Adv. Mater.* **32**, 2005228 (2020).

14 Li, Z. *et al.* Field-free topological behavior in the magnetic domain wall of ferrimagnetic GdFeCo. *Nat. Commun.* **12**, 5604 (2021).

15 Nagase, T. *et al.* Observation of domain wall bimerons in chiral magnets. *Nat. Commun.* **12**, 3490 (2021).

16 Xu, J. *et al.* Spontaneous Topological (Anti)meron Chains in the Domain Walls of Centrosymmetric Rare-Earth Magnet. *Adv. Funct. Mater.* **34**, 2314127 (2024).

17 Brüning, R. *et al.* Topological Meron-Antimeron Domain Walls and Skyrmions in a Low-Symmetry System. *Phys. Rev. X.* **15**, 021041 (2025).

18 Yoshimori, A. A New Type of Antiferromagnetic Structure in the Rutile Type Crystal. *J. Phys. Soc. Jpn.* **14**, 807-821 (1959).

19 Moriya, T. Anisotropic Superexchange Interaction and Weak Ferromagnetism. *Phys. Rev.* **120**, 91-98 (1960).

20 Koehler, W. C., Cable, J. W., Wilkinson, M. K. & Wollan, E. O. Magnetic Structures of Holmium. I. The Virgin State. *Phys. Rev.* **151**, 414-424 (1966).

21 Bogdanov, A. N. & Yablonskii, D. A. Thermodynamically stable "vortices" in magnetically ordered crystals. The mixed state of magnets. *Sov. Phys. JETP* **68**, 101 (1989).

22 Bogdanov, A. & Hubert, A. Thermodynamically stable magnetic vortex states in magnetic crystals. *J. Magn. Magn. Mater.* **138**, 255-269 (1994).

23 Bogdanov, A. N., Rößler, U. K., Wolf, M. & Müller, K. H. Magnetic structures and reorientation transitions in noncentrosymmetric uniaxial antiferromagnets. *Phys. Rev. B.* **66**, 214410 (2002).
24 Kimura, T. *et al.* Magnetic control of ferroelectric polarization. *Nature* **426**, 55-58 (2003).
25 Mostovoy, M. Ferroelectricity in Spiral Magnets. *Phys. Rev. Lett.* **96**, 067601 (2006).
26 Hubert, A. & Schäfer, R. *Magnetic Domains: The Analysis of Magnetic Microstructures*. (Springer Berlin Heidelberg, 2008).
27 Chakrabartty, D., Jamaluddin, S., Manna, S. K. & Nayak, A. K. Tunable room temperature magnetic skyrmions in centrosymmetric kagome magnet $Mn_4Ga_2Sn$. *Commun. Phys.* **5**, 189 (2022).
28 Chakrabartty, D. *et al.* Role of Competing Magnetic Exchange on Non-Collinear to Collinear Magnetic Ordering and Skyrmion Stabilization in Centrosymmetric Hexagonal Magnets. *ACS Nano* **19**, 3614-3623 (2025). https://doi.org/10.1021/acsnano.4c13876
29 Mühlbauer, S. *et al.* Skyrmion Lattice in a Chiral Magnet. *Science* **323**, 915-919 (2009).
30 Yu, X. Z. *et al.* Near room-temperature formation of a skyrmion crystal in thin-films of the helimagnet FeGe. *Nat. Mater.* **10**, 106 (2010).
31 Yu, X. Z. *et al.* Real-space observation of a two-dimensional skyrmion crystal. *Nature* **465**, 901-904 (2010).
32 Kezsmarki, I. *et al.* Neel-type skyrmion lattice with confined orientation in the polar magnetic semiconductor $GaV_4S_8$. *Nat. Mater.* **14**, 1116-1122 (2015).
33 Heinze, S. *et al.* Spontaneous atomic-scale magnetic skyrmion lattice in two dimensions. *Nat. Phys.* **7**, 713-718 (2011).
34 Seki, S., Yu, X. Z., Ishiwata, S. & Tokura, Y. Observation of skyrmions in a multiferroic material. *Science* **336**, 198-201 (2012).
35 Adams, T. *et al.* Long-Wavelength Helimagnetic Order and Skyrmion Lattice Phase in $Cu_2OSeO_3$. *Phys. Rev. Lett.* **108**, 237204 (2012).
36 Leonov, A. O. & Mostovoy, M. Multiply periodic states and isolated skyrmions in an anisotropic frustrated magnet. *Nat. Commun.* **6**, 8275 (2015).
37 Kurumaji, T. *et al.* N\'eel-Type Skyrmion Lattice in the Tetragonal Polar Magnet $VOSe_2O_5$. *Phys. Rev. Lett.* **119**, 237201 (2017).
38 Li, F., Nattermann, T. & Pokrovsky, V. L. Vortex Domain Walls in Helical Magnets. *Phys. Rev. Lett.* **108**, 107203 (2012).
39 Schoenherr, P. *et al.* Topological domain walls in helimagnets. *Nat. Phys.* **14**, 465-468 (2018).
40 Nayak, A. K. *et al.* Magnetic antiskyrmions above room temperature in tetragonal Heusler materials. *Nature* **548**, 561-566 (2017).
41 Jena, J. *et al.* Elliptical Bloch skyrmion chiral twins in an antiskyrmion system. *Nat. Commun.* **11**, 1115 (2020). https://doi.org/10.1038/s41467-020-14925-6
42 Ma, T. *et al.* Tunable Magnetic Antiskyrmion Size and Helical Period from Nanometers to Micrometers in a $D_{2d}$ Heusler Compound. *Adv. Mater.* **32**, 2002043 (2020).
43 Peng, L. *et al.* Controlled transformation of skyrmions and antiskyrmions in a non-centrosymmetric magnet. *Nat. Nanotechnol.* **15**, 181-186 (2020).
44 Meshcheriakova, O. *et al.* Large Noncollinearity and Spin Reorientation in the Novel $Mn_2RhSn$ Heusler Magnet. *Phys. Rev. Lett.* **113**, 087203 (2014).
45 Giri, B. *et al.* Robust topological Hall effect driven by tunable noncoplanar magnetic state in Mn-Pt-In inverse tetragonal Heusler alloys. *Phys. Rev. B.* **102**, 014449 (2020).
46 Rybakov, F. N. B., E. Excalibur software. (2020).

https://doi.org/http://quantumandclassical.com/excalibur/
47 Chovan, J., Papanicolaou, N. & Komineas, S. Intermediate phase in the spiral antiferromagnet $Ba_2CuGe_2O_7$. *Phys. Rev. B.* **65**, 064433 (2002).
48 Rybakov, F. N., Eriksson, O. & Kiselev, N. S. Topological invariants of vortices, merons, skyrmions, and their combinations in continuous and discrete systems. *Phys. Rev. B.* **111**, 134417 (2025).
49 Song, D. *et al.* Robust nature of the chiral spin helix in $CrNb_3S_6$ nanostructures studied by off-axis electron holography. *Phys. Rev. B.* **102**, 064432 (2020).
50 Zhang, X., Zhou, Y. & Ezawa, M. Antiferromagnetic skyrmion: stability, creation and manipulation. *Scientific reports* **6**, 1-8 (2016).
51 Barker, J. & Tretiakov, O. A. Static and Dynamical Properties of Antiferromagnetic Skyrmions in the Presence of Applied Current and Temperature. *Phy. Rev. Lett.* **116**, 147203 (2016).
52 Song, D. *et al.* Steady motion of 80-nm-size skyrmions in a 100-nm-wide track. *Nat. Commun.* **15**, 5614 (2024).
53 Müller, J. *et al.* Magnetic Skyrmions and Skyrmion Clusters in the Helical Phase of $Cu_2OSeO_3$. *Phys. Rev. Lett.* **119**, 137201 (2017).
54 He, Z. *et al.* Experimental observation of current-driven antiskyrmion sliding in stripe domains. *Nat. Mater.* **23**, 1048-1054 (2024).
55 Karube, K. *et al.* Room-temperature antiskyrmions and sawtooth surface textures in a non-centrosymmetric magnet with $S_4$ symmetry. *Nat. Mater.* **20**, 335-340 (2021).
56 Karube, K. *et al.* Doping Control of Magnetic Anisotropy for Stable Antiskyrmion Formation in Schreibersite (Fe,Ni)3P with S4 symmetry. *Adv. Mater.* **34**, 2108770 (2022).
57 Yu, X. Z. *et al.* Transformation between meron and skyrmion topological spin textures in a chiral magnet. *Nature* **564**, 95-98 (2018).
58 Rikken, G. L. J. A., Fölling, J. & Wyder, P. Electrical Magnetochiral Anisotropy. *Phys. Rev. Lett.* **87**, 236602 (2001).
59 Fontana, P. *et al.* Quantum Geometry and the Electric Magnetochiral Anisotropy in Noncentrosymmetric Polar Media. *Phys. Rev. Lett.* **135**, 106602 (2025).

## Methods

### Micromagnetic Simulations:

Micromagnetic simulations that include the dipole-dipole interaction are carried out using the *Excalibur* code[46]. The DMI term in (1) is defined by combinations of Lifshitz invariants:

$$\Lambda_{ij}^{(k)}(m) = m_i \cdot \frac{\partial m_j}{\partial r_k} - m_j \cdot \frac{\partial m_i}{\partial r_k}$$

The parameters $A$ and $D$ are the exchange stiffness and DMI constant, respectively. To calculate the Lorentz TEM image of the simulated spin textures, we have used a complete 3D micromagnetic model including the dipole-dipole interaction energy term. For these simulations the following parameters were used: exchange stiffness constant, $A = 6 \times 10^{-12}$ J/m, DMI constant $D = -0.6 \times 10^{-3}$ J/m$^2$ (due to symmetry, the sign of this constant alternates when the coordinate system is rotated by 90° around the Z axis), and saturation magnetization $M_s = 200 \times 10^3$ A/m. The sample was discretized into cuboids of size 5 nm$^3$ each. The total size of the simulated domain is 256 × 150 × 38 cuboids with periodic boundary conditions along $x$ and $y$. To calculate the theoretical Lorentz TEM image, we use the accelerated voltage (electron energy) $U$ = 200 kV, and the defocus distance $\Delta z$ = 1000 $\mu$m from our experiments.

### Sample preparation and characterizations:

Polycrystalline samples of $Mn_{1.4}Pt_{0.9}Pd_{0.1}Sn_{1-x}In_x$ were prepared by arc melting the stoichiometric amounts of high-purity constituent elements. The samples were kept at 750˚C for seven days after being sealed in high vacuum quartz tubes. The structural phase of the samples was confirmed by room temperature powder X-ray diffraction (XRD) using Cu-$K_\alpha$ radiation (Rigaku). The composition and homogeneity of the samples were checked in field emission scanning electron microscopy (FESEM) and energy dispersive X-ray spectroscopy (EDS). To check the composition homogeneity in the = 0.4 lamellae, we performed EDAX in a JEOL JEM200 Transmission Electron Microscope.

**Magnetic, Transport and Neutron Diffraction measurements:**

DC and AC magnetic measurements were carried out using a superconducting quantum interference device vibrating sample magnetometer (SQUID VSM) (MPMS-3). For the electrical transport study, first, the grain of interest was determined by electron back scattered diffraction (EBSD). Several micro-devices in Hall bar geometry were prepared using standard lift off method in FIB (TESCAN GAIA). The electrical contact was made with sample to gold electrode by depositing Pt in FIB. To reduce the contact resistance a layer of Ti(10nm)/Au (80 nm) deposited on the devices using the standard photolithography technique. To remove the gold layer from the unwanted area of the device we used low energy (5 kV, 15 pA) ion beam which does not alter the physical properties of the sample. The typical size of the devices is approximately $7 \times 4 \times 1 \mu m^3$. For neutron diffraction (ND) measurements, we prepared polycrystalline samples of $Mn_{1.4}PtSn_{1-x}In_x$ with x=0.2, 0.4, and x=0.6. The powder neutron diffraction data were taken using the PD1 neutron diffractometer at the Dhruva reactor in Bhabha Atomic Research Center, Mumbai, India.

**LTEM measurements:**

For the real space visualization of magnetic textures using Lorentz Transmission Electron Microscopy (LTEM) imaging, thin lamellae of $Mn_{1.4}Pt_{0.9}Pd_{0.1}Sn_{1-x}In_x$ were cut from the polycrystalline sample using a 30 kV Ga ion-based Focused Ion Beam (FIB) (ZEISS). Prior to the lamella preparation, we utilized electron back scattered diffraction (EBSD) to find suitable grains for [001] oriented lamellae preparation from the polycrystalline samples. A very low dose factor was used to thin down the lamellae during the final stage. The (001) orientations of the lamellae were confirmed by performing selected area electron diffraction (SAED) in a 200 kV JEOL JEM200 microscope. The thickness of the lamella that used for the LTEM measurement for the $x$ =0.4 sample is approximately 150 nm. The real space imaging of the magnetic contrast was observed using the LTEM mode in JEOL 200 kV NEOARM and JEOL F200 microscopes.

**Transport of intensity equation (TIE):**

LTEM imaging is a versatile technique to visualize various nano to micro scale magnetic textures in thin lamella samples. Interaction of the electrons with the in-plane magnetic field of the sample results in the convergence(divergence) of the electron beam that produces bright(dark) contrast under different defocus conditions. If a magnetic image appears as bright in over focus (OF) condition, then it changes to dark in the under focus (UF) and vice versa. Extraction of the actual spin texture is possible by solving the transport of intensity equation (TIE) from the intensity map of the series of digital images. The TIE solves the differential equation between intensity and electron's phase given as-

$$\frac{2\pi}{\lambda}\frac{\partial^2 I(x,y,z)}{\partial^2 z} = -\nabla_{xy} \cdot [I\,(x, y, z)\,\nabla_{xy}\phi(x, y, z)], \qquad (2)$$

where, $\lambda$ is electron wavelength, $I$ is the intensity of electrons beam, and $\phi$ is the phase of electrons. A series of systematic images can be acquired to map the in-plane magnetization distribution of the thin lamella. For example, an equal defocus length of UF, OF, along with in-focus (IF) images can be used to map out the magnetization distribution. First, we extract the electron-phase image and then mapped out the magnetization textures using commercially available QPt software.

**Electron Holography measurements:**

Off-axis electron holography was performed using an electrostatic biprism placed in a conjugate image plane of an FEI Titan transmission electron microscope operated at an accelerating voltage of 300 kV. The microscope was run in aberration-corrected Lorentz mode to ensure field-free conditions at the specimen position. Out–of plane magnetic field was applied using the objective lens. Sample temperature was varied between 95 and 380 K using a liquid nitrogen cooled specimen holder (Gatan model 636). Lorentz images and off-axis electron holograms were acquired using both a 2k × 2k CCD camera (Gatan Ultrascan) and a direct electron detection K2-IS camera (Gatan). To enhance the signal-to-noise ratio of the reconstructed phase images, typically 12 object holograms and 12 vacuum reference holograms were recorded with an exposure time of 4 seconds each. The holograms were processed and reconstructed using a standard fast Fourier transform–based method implemented in Holoworks 6.0 software (Gatan).

**Data availability**

All data used to obtain the conclusions in this work are presented in the paper and/or the Supplementary Materials. Other data may be requested from the authors. Please direct all inquiries to A.K.N. (ajaya@niser.ac.in).

**Acknowledgments**

A.K.N. acknowledges the Department of Atomic Energy (DAE) and SERB research grant (CRG/2022/002517) of the Government of India for financial support. A.K.N. and S.S.P.P. acknowledge the Max Plank Society for support under the Max Planck-India partner group project. A.K.N. and B.G. thanks Amitabh Das, BARC, Mumbai for the neutron diffraction measurement. This research was partially funded by the Deutsche Forschungsgemeinschaft (DFG, German Research Foundation) through project number 49254781 (TRR 360). F.N.R. acknowledges support from the Swedish Research Council (grant 2023-04899).

**Author contributions**

A.K.N. conceived the idea of the present work and led the project. B.G. prepared and characterized polycrystalline samples, prepared the TEM lamellae, and analyzed the neutron diffraction data. F.N.R. and N.S.K. carried out the micro-magnetic simulations, topological analysis and wrote a part of the manuscript. B.G., S.P.S., and R.S. carried out the magnetic and ac susceptibility measurements. T.K., B.G., D.C., and A.K.N. performed the LTEM measurements with help of F.B. and I.K. The micro-devices for Hall transport study were prepared by B.G. with the help from A.K.S., A.P., and B.P. The electron holography study was performed by A.H.T. and S.P.S. with the input from R.E.D. All authors contribute to the data analysis. A.K.N. and B.G. wrote the manuscript with substantial contributions from N.S.K., F.N.R., F.B., I.K., and S.S.P.P.

## Additional information

Supplementary Information for the present manuscript is available at https:

**Competing interests:** The authors declare that they have no competing interests.

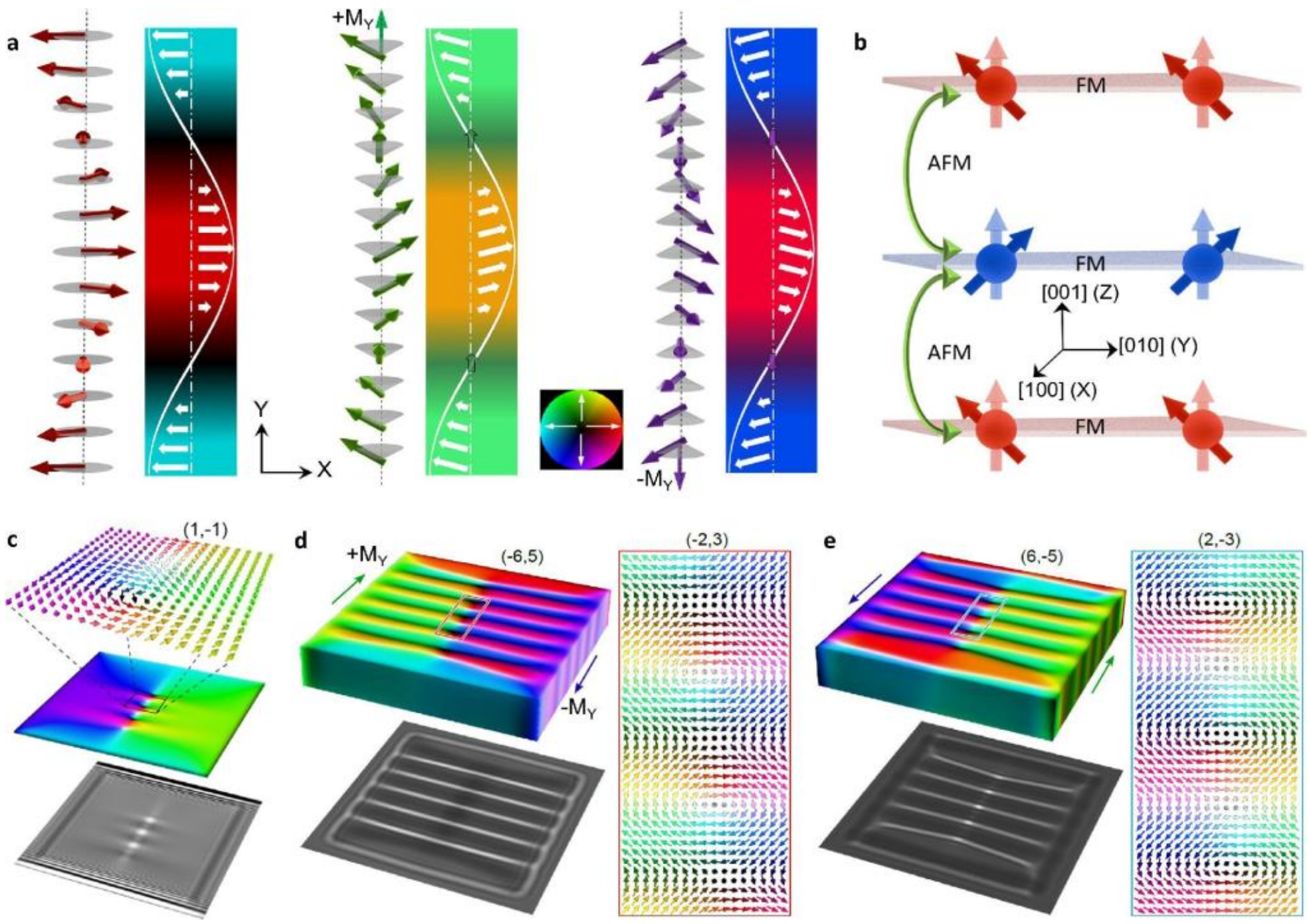

Figure 1: **Spin-spirals and their role in bimeron formation in a system with $D_{2d}$ symmetry. a** Schematics of a right-handed spin helix (left panel), and right-handed conical helices with net magnetization pointing up/down along the helical axis (middle/right panel). For each case the expected variation of the magnetization profile (according to the color wheel between the middle and right panels) is additionally shown on the right side. **b** Schematic showing intra-plane ferromagnetic (FM) and inter-plane antiferromagnetic (AFM) interactions in two different magnetic sublattices represented by magnetic moments with red and blue arrows. Semi-transparent arrows represent collinear FM ordering state, whereas the solid arrows refer to canted AFM ordering between the two magnetic sublattices. **c** Bimeron configuration with topological charge (1, −1) formed by alternating merons with topological indices (0, −1) and (1, 0) in a quasi 2D sample of thickness 5 nm. **d** and **e** Analogous configurations composed of alternating merons in a finite thickness sample of 185 nm, where the domains exhibit a cone phase with opposite net magnetizations indicated by the arrows. The right panels of **d** and **e** show the spin configuration in the central plane of the sample and correspond to the zoomed-in regions marked by red and blue rectangles in the main panels. The bottom panels in **c**, **d**, and **e** present corresponding over-focused Lorentz TEM images simulated for a defocus distance of 1000 $\mu$m and an accelerating voltage of 200 keV.

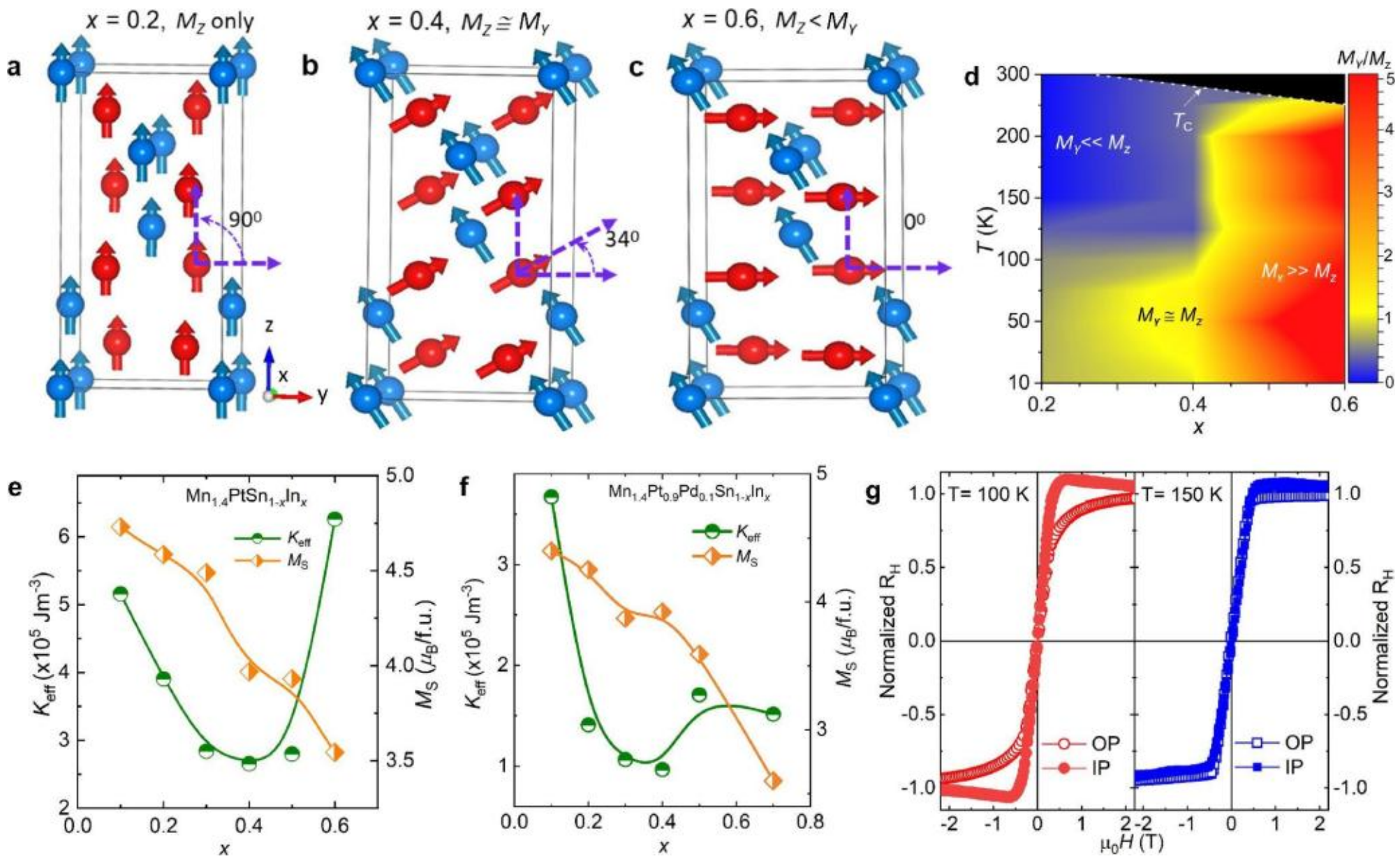

Figure 2: **Anisotropy engineering in Mn-Pt(Pd)-Sn-In based $D_{2d}$ symmetric materials.** **a-c** Evolution of out of plane (OP) magnetization component $M_Z$ and in plane (IP) magnetization $M_Y$ with indium doping in $Mn_{1.4}PtSn_{1-x}In_x$, obtained from powder neutron diffraction experiments. The blue and red balls correspond to the Mn atoms at 4a and 8d Wyckoff positions, respectively. **d** Temperature de-pendent $M_Y$ /$M_Z$ for different In concentrations obtained from the neutron diffraction study. The scale bar shows the $M_X$/$M_Z$ magnitudes. Variation of effective anisotropy ($K_{eff}$) and saturation magnetization $M_S$ with In concentration ($x$) for **e** $Mn_{1.4}PtSn_{1-x}In_x$ and **f** $Mn_{1.4}Pt_{0.9}Pd_{0.1}Sn_{1-x}In_x$. **g** Direction dependent normalized Hall resistivity data taken at 100 K and 150 K for $Mn_{1.4}Pt_{0.9}Pd_{0.1}Sn_{0.6}In_{0.4}$.

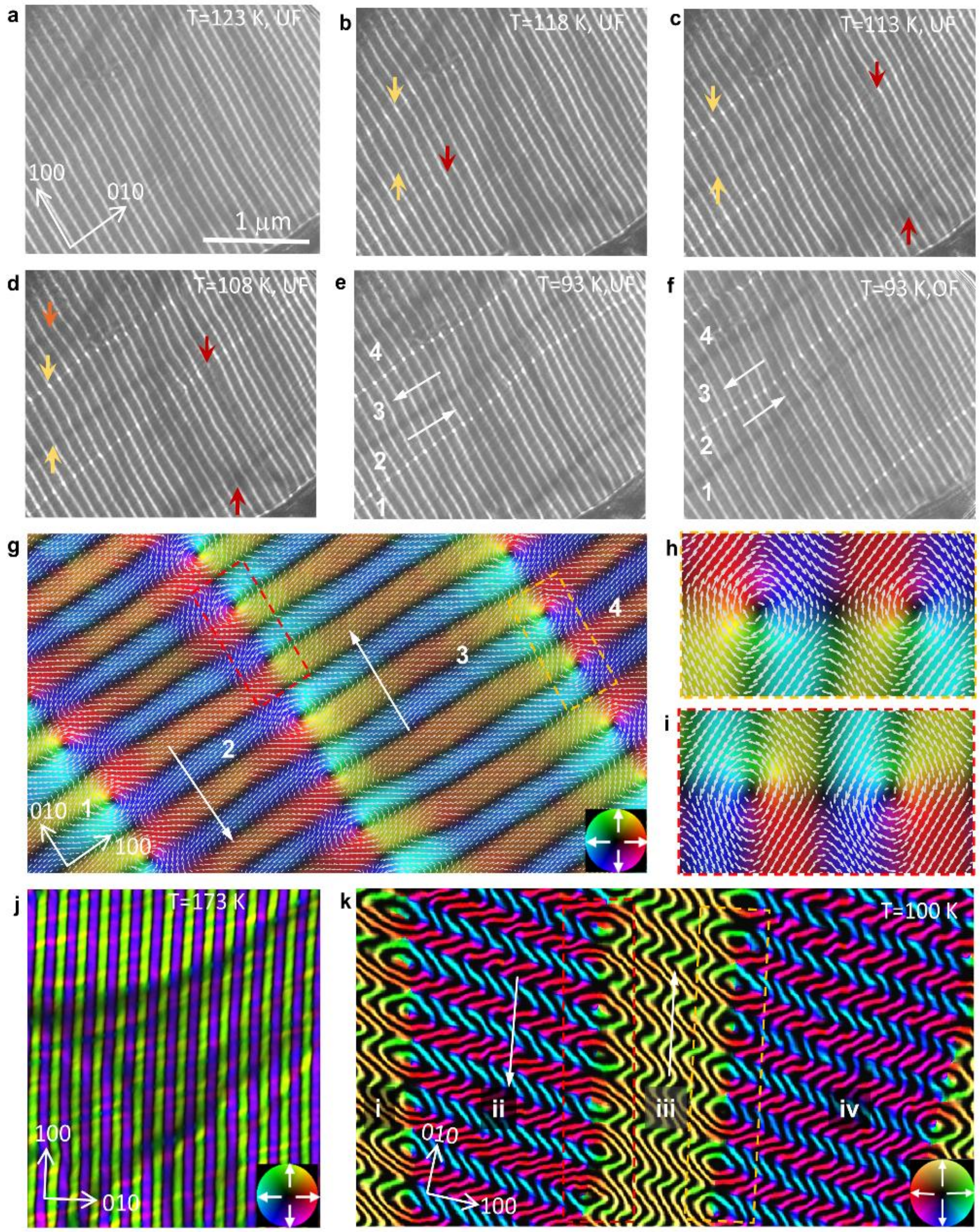

Figure 3: **Temperature evolution of merons in $\mathbf{Mn_{1.4}Pt_{0.9}Pd_{0.1}Sn_{0.6}In_{0.4}}$. a-e** Zero-field under-focused LTEM images taken at different temperatures. **f** Over-focused (OF) LTEM image at 93 K. The vertical arrows points to the appearance of additional bright and dark dotted chain-like patterns on top of the magnetic stripes. In **a-f**, additional evenly spaced faint dark lines running from the top-right to the bottom-left correspond to the curtaining effect caused by the FIB milling. In **a-f**, additional vertical dark contrast corresponds to diffraction bending contour. **g** TIE analysis of the LTEM images taken at 93 K. Different colors represents the directional variation

of the local magnetic field, as shown by the color wheel. **h** and **i** show enlarged views of the magnetic profile taken from the area enclosed by the yellow and red dashed boxes in **g**, respectively. **j** TIE analysis of the conventional helical-like patterns at 173 K. **k** Magnetic induction maps of the stripe-magnetic patterns and bimeron-chains recorded using off-axis electron holography. In **g** and **k** different domains are numbered as i, ii, iii and iv. The direction of staggered magnetic moment pointing along or opposite to the [010] is shown by the long white arrows in domains 2 and 3 in **e, f, g and k**. The ground-state magnetic patterns observed by LTEM and off-axis electron holography were obtained from two independent sets of measurements.

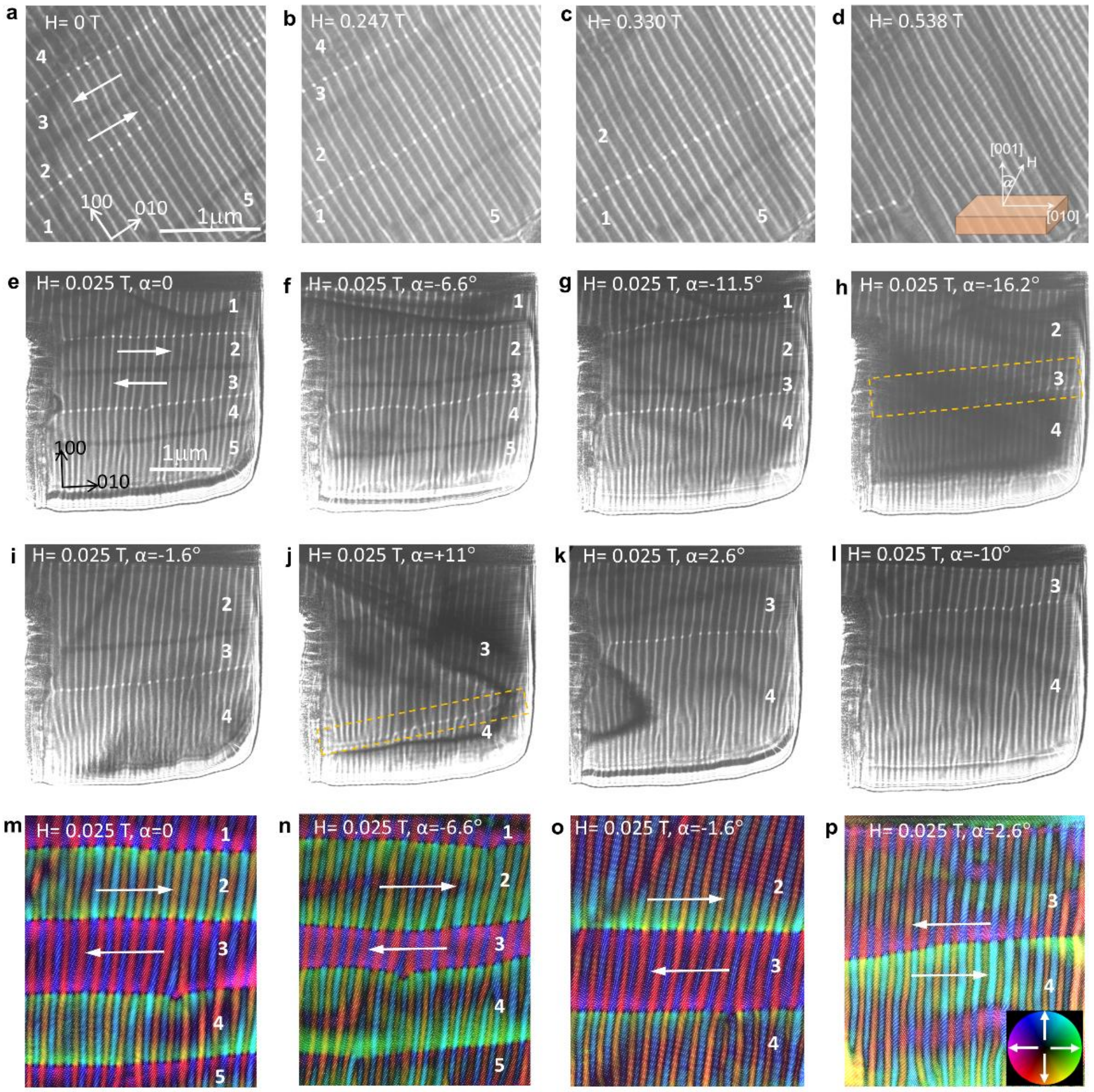

Figure 4: **Confined bimeron motion along the stripes. a-d** Evolution of UF LTEM images taken at different magnetic fields at $T = 93$ K for $Mn_{1.4}Pt_{0.9}Pd_{0.1}Sn_{0.6}In_{0.4}$. The inset in **d** shows a schematic depicting the direction of applied magnetic field. The lamella was tilted by an angle of $\alpha$ = 4° towards [010] to apply a small magnetic field along the propagation direction of the conical-helix. In **a-d**, additional evenly spaced faint dark lines running from the top-right to the bottom-left correspond to the curtaining effect caused by the FIB milling. **e-l** OF LTEM images taken at $T$ = 93 K with different tilting angles. First, an out-of-plane magnetic field of 25 mT was applied with the sample at zero tilt. The sample was then tilted back and forth to introduce an in-plane magnetic field component, enabling the stabilization and controlled motion of a single bimeron chain, as shown in **j–l**. In **e–l**, additional dark contrast, together with the nearly horizontal bright/dark dotted contrast associated with the bimeron chains, arises from diffraction bending contours. **m-p** TIE analysis of the magnetic patterns recorded at $\alpha$ = 0°, -6.6°, -1.6, and +2.6. Different domains are numbered with 1, 2, 3 and 4. The direction of staggered magnetic moment pointing along or opposite to the [010] is shown by the long white arrows. The additional color contrast observed within some domains in **m–p** arises from diffraction contrast superimposed on the magnetic contrast. The magnetic field shown in the figure represents the total applied field.

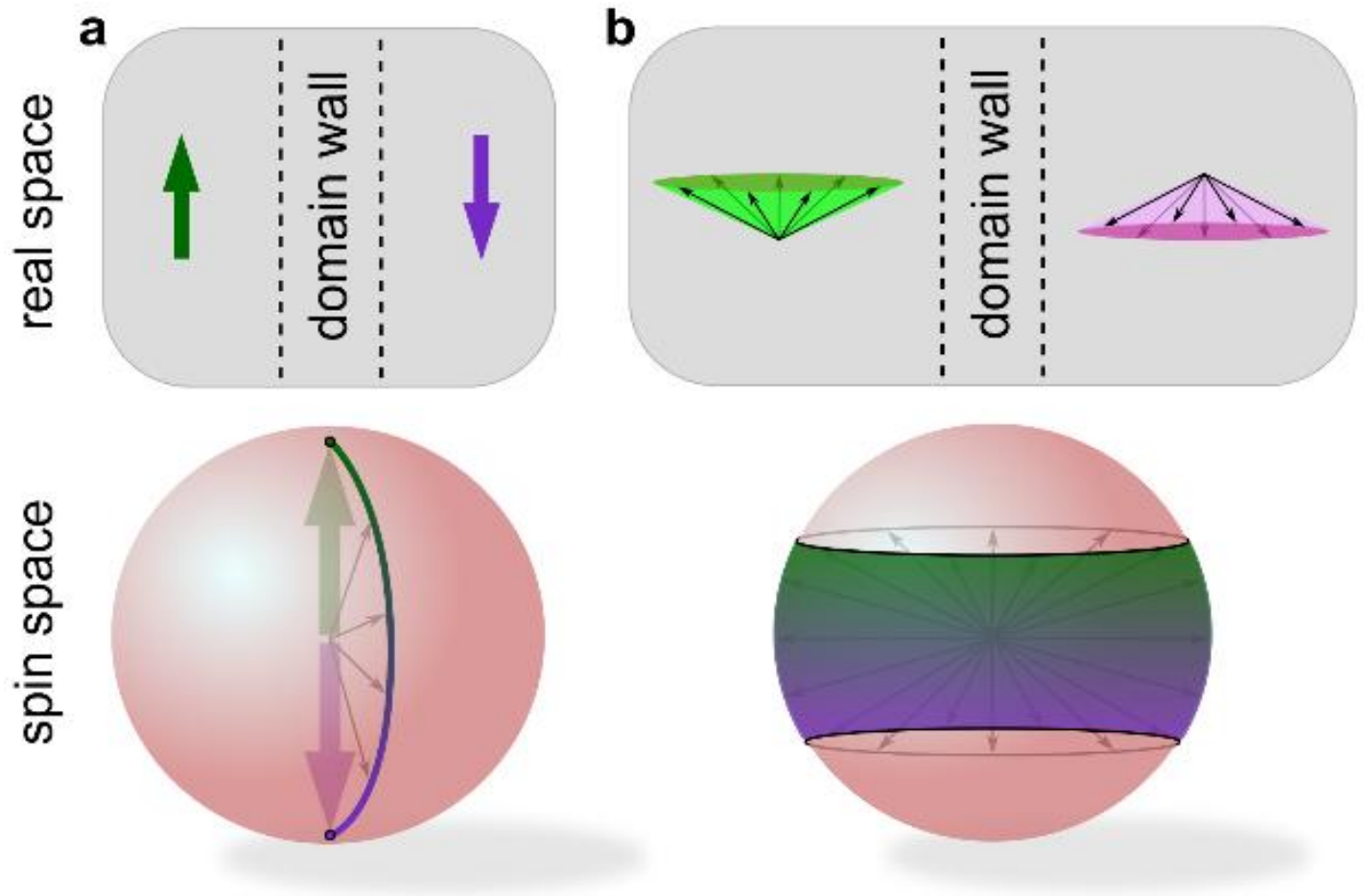

Figure 5: **Minimal models of domain walls**. **a** Domain wall separating two ferromagnetic domains and composed of spins form a sphere arc. **b** Domain wall separating two conical domains and composed of spins form a spherical zone.